\documentclass{JAC2003}
\pdfoutput=1
\usepackage{mathptmx}
\usepackage{graphicx}
\usepackage{booktabs}
\usepackage[english]{babel}
\usepackage[tracking,spacing,kerning,babel,letterspace=20]{microtype}

\makeatletter
\renewcommand{\section}{\@startsection {section}{1}{0mm}{9pt}{3pt}{\centering\scshape\bfseries\fontsize{12}{15}\selectfont\MakeUppercase}}
\renewcommand{\subsection}{\@startsection {subsection}{2}{0mm}{6pt}{3pt}{\raggedright\itshape\fontsize{12}{15}\selectfont}}
\makeatother

\begin{document}
\title{\textsc{GAS ELECTRON MULTIPLIERS}\linebreak \textsc{VERSUS MULTIWIRE PROPORTIONAL CHAMBERS}}

\author{Serge Duarte Pinto\thanks{S.C.DuartePinto@tudelft.nl}, TU Delft, Netherlands\\
Jens Spanggaard, \textsc{cern}, Geneva Switzerland\\}

\maketitle

\begin{abstract}
Gas Electron Multiplication technology is finding more and more applications in beam instrumentation and at \textsc{cern} these detectors have recently been adapted for use in transverse profile measurements at several of our facilities.
In the experimental areas of \textsc{cern}'s Antiproton Decelerator, low energy Gas Electron Multipliers successfully replaced all Multi-Wire Proportional Chambers in 2012 and another detector type has now been developed for high energy applications in the experimental areas of the \textsc{sps}, totaling a potential of more than a hundred profile detectors to be replaced by \textsc{gem} detectors of different types.
This paper aims to describe the historical evolution of \textsc{gem} technology by covering the many different applications but with specific focus on its potential to replace Multi-Wire Proportional Chambers for standard transverse profile measurement.
\end{abstract}

\section{Introduction}
The Gas Electron Multiplier (\textsc{gem}) was invented at \textsc{cern} in 1997 by Fabio Sauli \cite{GEM}.
Although it was originally introduced as a preamplifier to help the then novel microstrip gas chambers cope with the high particle rate in the \textsc{hera-b} experiment at \textsc{desy} \cite{HeraB}, the \textsc{gem} soon became the basis of a detector in its own right.
Once it was recognized that \textsc{gem}-based detectors had many attractive features besides the demonstrated superb rate capability, these detectors also gained ground in other fields than high-energy physics experiments \cite{MPLA}; including notably in beam instrumentation.

In the experimental areas of \textsc{cern}, \textsc{gem} detectors have been developed in the past few years to replace multiwire proportional chambers (\textsc{mwpc}) as transverse profile monitors.
At the beam lines of the Antiproton Decelerator (\textsc{ad}) all wire chambers have been replaced by single \textsc{gem} detectors of a very lightweight design \cite{Kondo,IPAC10,DIPAC11,jinst}, to avoid affecting the measured profile through interaction of the low energy beam (5.3 MeV) with the detector.
Prototype triple \textsc{gem} detectors have also been tested at various high energy beam lines; these detectors are of a different design and sometimes different size, as will be discussed below.

\section{The technology}
The working principle of \textsc{mwpc}s and \textsc{gem}-based detectors is very similar.
\begin{figure}
\includegraphics[width=\columnwidth]{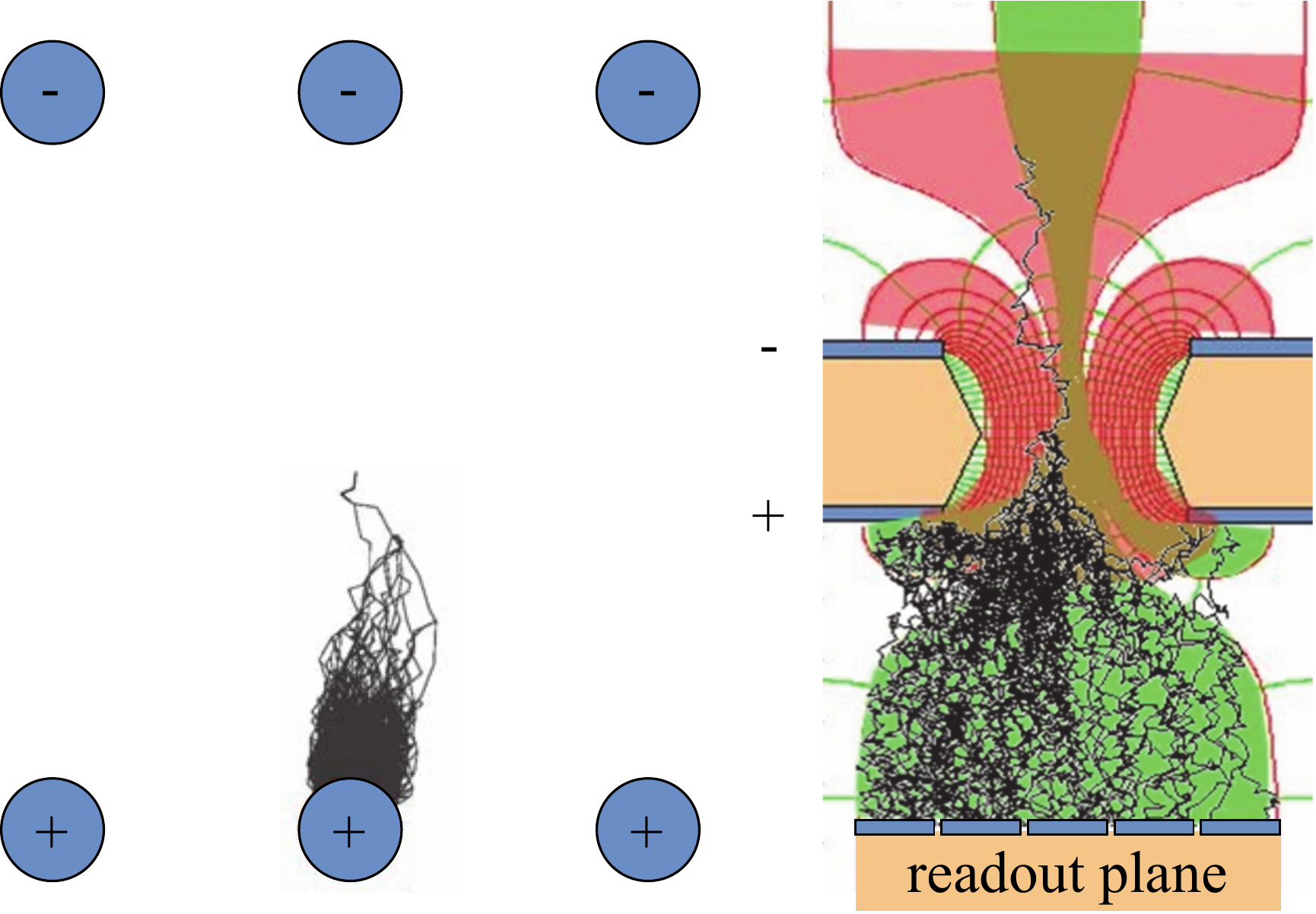}
\caption{Working principle of \textsc{mwpc} (left) and \textsc{gem} (right) illustrated. Electron avalanches as simulated by Garfield\protect\footnotemark\ are shown for both technologies; black paths are electron trajectories, the drift of ions is not indicated.}
\label{MWPCvsGEM}
\end{figure}
The ionization electrons liberated by passage of a charged particle through a gas medium are attracted by a low electric field ($<10$ keV) to a region with a higher field.
This higher field (typically in the range of 10--100 keV, strongly depending on gas mixture and pressure)\footnotetext{A Monte Carlo simulation program developed at \textsc{cern}. Author: Rob Veenhof (http://garfield.web.cern.ch/garfield/)} causes a proportional mode gas avalanche, in which the electrons cause further ionization, thereby exponentially increasing the total charge.
As all these charges (electrons and ions) drift towards electrodes where they are collected, their motion induces a signal on readout electrodes.

Figure~\ref{MWPCvsGEM} illustrates how this is implemented in the case of wire chambers and \textsc{gem}s.
The strong field region where multiplication takes place in a wire chamber is the $1/r$ field close to an anode wire, which also collects the electrons, and often serves as a readout electrode.
In a \textsc{gem} foil, the strong field is formed in millions of microscopic holes in a dielectric foil by biasing the top and bottom electrodes.
The figure shows a computed field line pattern in a \textsc{gem} hole, strongly  focusing in the center of the hole.

There are some intrinsic differences between these two gas detectors.
In a \textsc{gem} detector, the \textsc{gem} foil screens the movement of ions (above the \textsc{gem}) from the readout elements (below), thereby eliminating the characteristic \emph{ion tails} in the signals of \textsc{mwpc}s.
In wire chambers, ions generated in the avalanche remain in the gas for tens or hundreds of microseconds, reducing the field strength around the anode wires (and therefore the gas gain).
The ions generated in a \textsc{gem} avalanche leave the hole within $\sim100$ ns, which explains the orders of magnitude higher rate capability of \textsc{gem}s.

\begin{figure*}
\includegraphics[width=\textwidth]{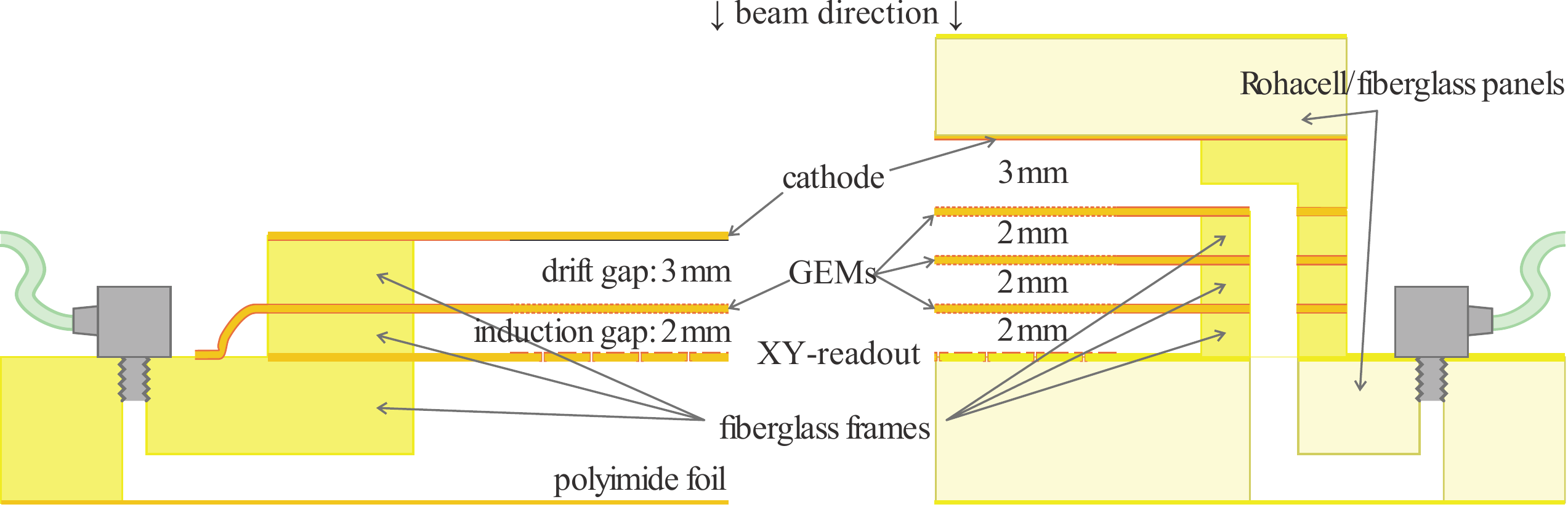}
\caption{Schematic view of how some of our detectors are built up. Left, a single \textsc{gem} detector with a light entrance window, as installed in the Antiproton Decelerator. Right, a triple \textsc{gem} detector made robust by light and stiff fiberglass/foam panels on front and back.}
\label{SingleTriple}
\end{figure*}

Wire chambers also suffer from aging.
The mechanisms and failure modes depend on the radiation environment and gas mixture.
The avalanche plasma tends to introduce free radicals in the gas, and deposit various sorts of chemical compounds on the wire surface.
The results are often non-uniformity of gain over the sensitive area and broken wires.
\textsc{Gem}s have proven to be much less prone to aging, and their limits of integrated charge still need to be found.

An advantage of wire chambers that may be of critical importance for some applications is the fact that they can be designed such that they present very little material to the beam.
The delay wire chambers \cite{DWC} in use in the \textsc{cern} experimental areas have only two thin polymer films, even thinner metal wire planes and a few millimeters of gas in the beam, resulting in a material budget of $\sim0.03\%X_0$.
By contrast, the lightest triple \textsc{gem} we could build would still present at least $0.15\%$ of $X_0$.

\section{Developments}
We will briefly describe some of the recent \textsc{gem} detector developments.
More details on the \textsc{ad} beam profile monitor can be found in \cite{IPAC10,DIPAC11,jinst}.

\subsection{Low energy antiprotons}
The wire chambers that were previously used at the Antiproton Decelerator to measure transverse beam profiles consisted of separate chambers to measure the horizontal (upstream) and vertical (downstream) profiles.
The 5.3 MeV antiproton beam was entirely absorbed by the upstream chamber, so that the downstream chamber only measured the profile of annihilation products.
The \textsc{gem}-based successors to these wire chambers are designed to read out both profiles in one plane, by a readout board that shares charge equally between horizontal and vertical elements.
This detector also has a very transparent polyimide entrance window, with a $\sim100$ nm chromium layer used as a cathode to bias the drift gap above the \textsc{gem} foil.
This ensures that almost all beam particles enter the detector, cause ionization in the drift gap, and then mostly stop and annihilate in the \textsc{gem} foil and the readout board.
The left hand side of figure~\ref{SingleTriple} shows this schematically.

\begin{figure}[b]
\includegraphics[width=.599\columnwidth]{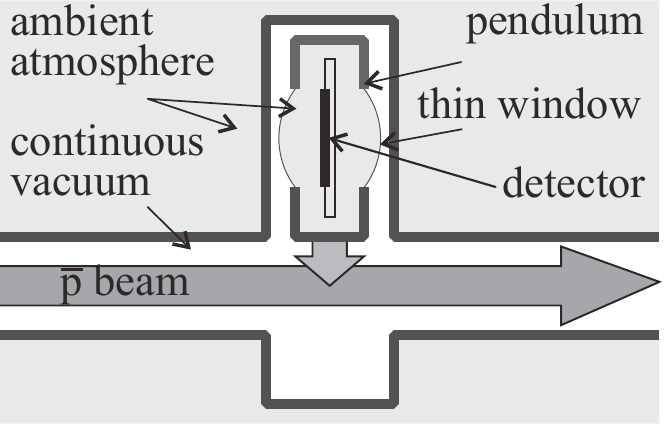}\includegraphics[width=.4\columnwidth]{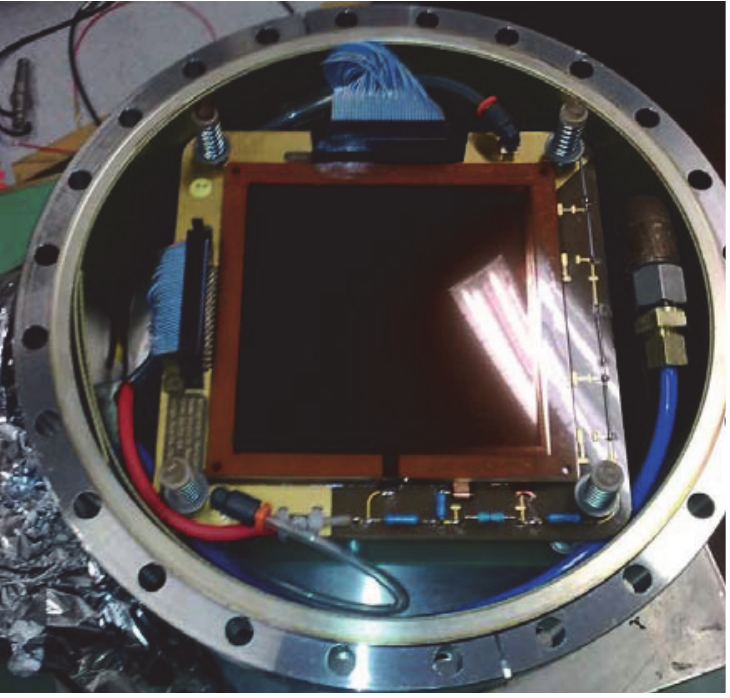}
\caption{Left, a schematic view of how a profile detector is situated in a pendulum that can move through teh beam vacu\-um. Right, a single \textsc{gem} detector installed in a pendulum, just before closing the lid.}
\label{pendulum}
\end{figure}

The low beam energy of the \textsc{ad} demands that the beam vacuum is uninterrupted during normal operation.
Only during machine development are detectors put in the beam for a destructive profile measurement.
To this end, the detectors are installed in a pendulum that can swing in or out of the beam, with the inside of the pendulum in contact with ambient atmosphere.
Figure~\ref{pendulum} shows how a detector is installed in a pendulum.

\begin{figure*}
\includegraphics[width=\textwidth]{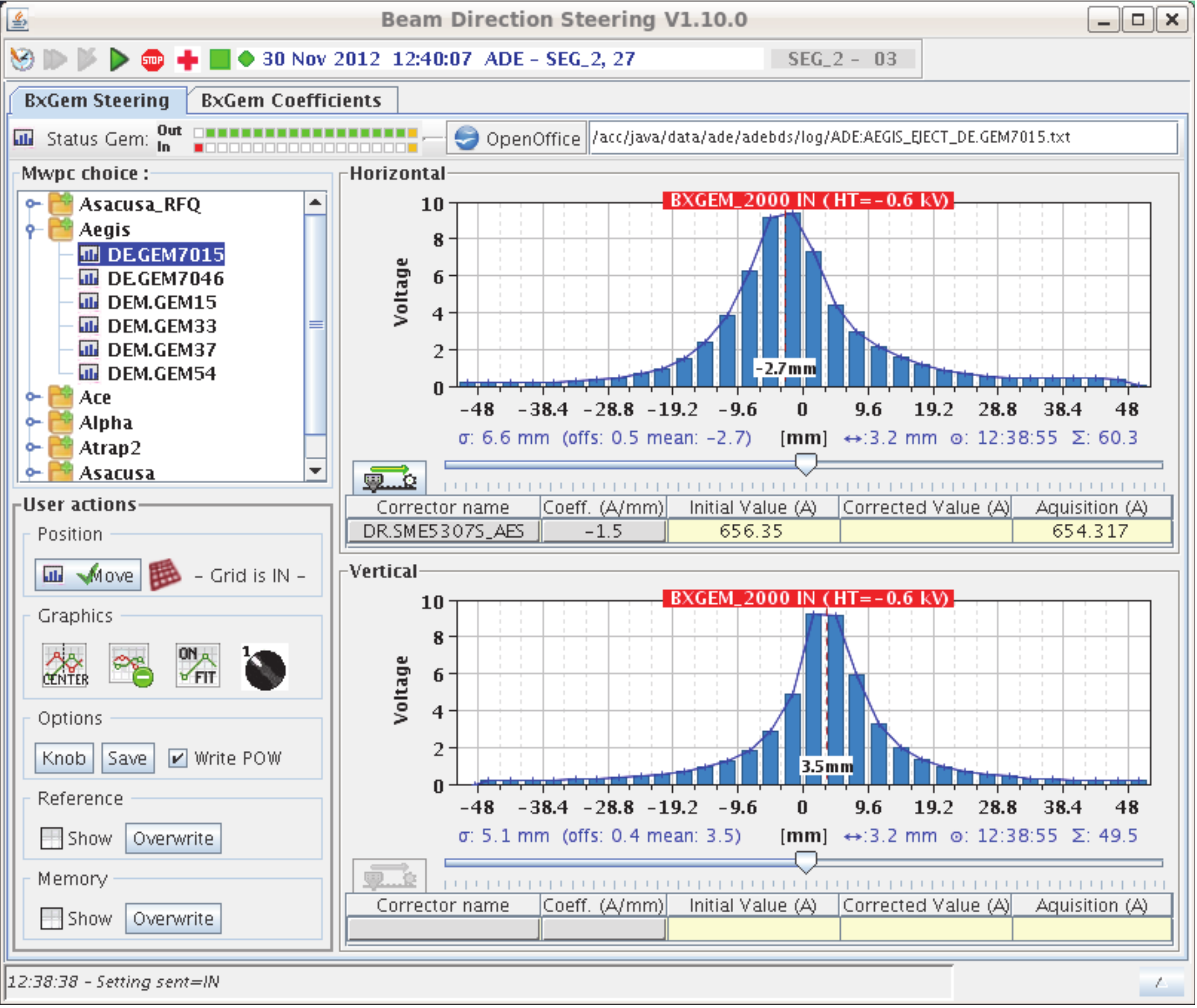}
\caption{A screenshot from the graphical user interface from which the \textsc{ad} profile monitors are operated and read out.}
\label{GUI}
\end{figure*}

These detectors have now been installed throughout the \textsc{ad}.
They have delivered profiles reliably, without the severe distortions common with the wire chambers.
The shape of the profile does not depend on particle rate or gas gain, indicating that the detector itself does not introduce distortions.
In addition, the integrated charge of the horizontal profile matches that of the vertical profile, which shows that the sharing of charge between readout elements is working as predicted.
The ionization from the 5.3 MeV beam is so strong that even the amplification of a single \textsc{gem} is not strictly needed.
We have demonstrated that at the nominal \textsc{ad} intensity ($\sim3\cdot10^7\ \bar{\textrm{p}}$ per spill) the detector could be operated as an ionization chamber, with just a cathode and a readout board, yielding equally good profiles.
But since the \textsc{ad} occasionally runs at a higher energy of 126 MeV for the \textsc{ace} experiment, a \textsc{gem} is included to provide about three orders of magnitude dynamic range.
Figure~\ref{GUI} shows a screenshot of the graphical user interface \textsc{ad} operators use to control and read out the monitors.

\subsection{High energy beams}
The high energy beam lines at \textsc{cern} use various kinds of \textsc{mwpc}s for transverse profiling and spectroscopy downstream of the production targets.
Most detectors have an active area of $10\times10$ cm$^2$, with some larger $20\times20$ cm$^2$ detectors on the M2 muon beam line used by the \textsc{compass} experiment.

\begin{figure}[b]
\includegraphics[width=\columnwidth]{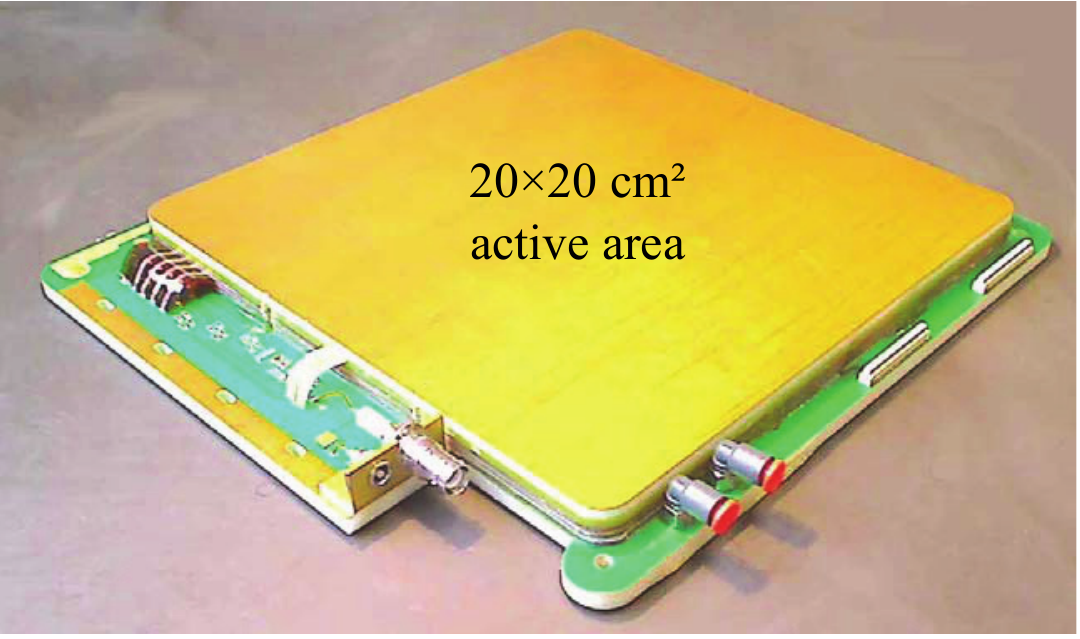}
\caption{Prototype of a $20\times20$ cm$^2$ triple \textsc{gem} detector, with stiff foam panels on the front and back.}
\label{BigGEM}
\end{figure}

\begin{figure}
\includegraphics[width=\columnwidth]{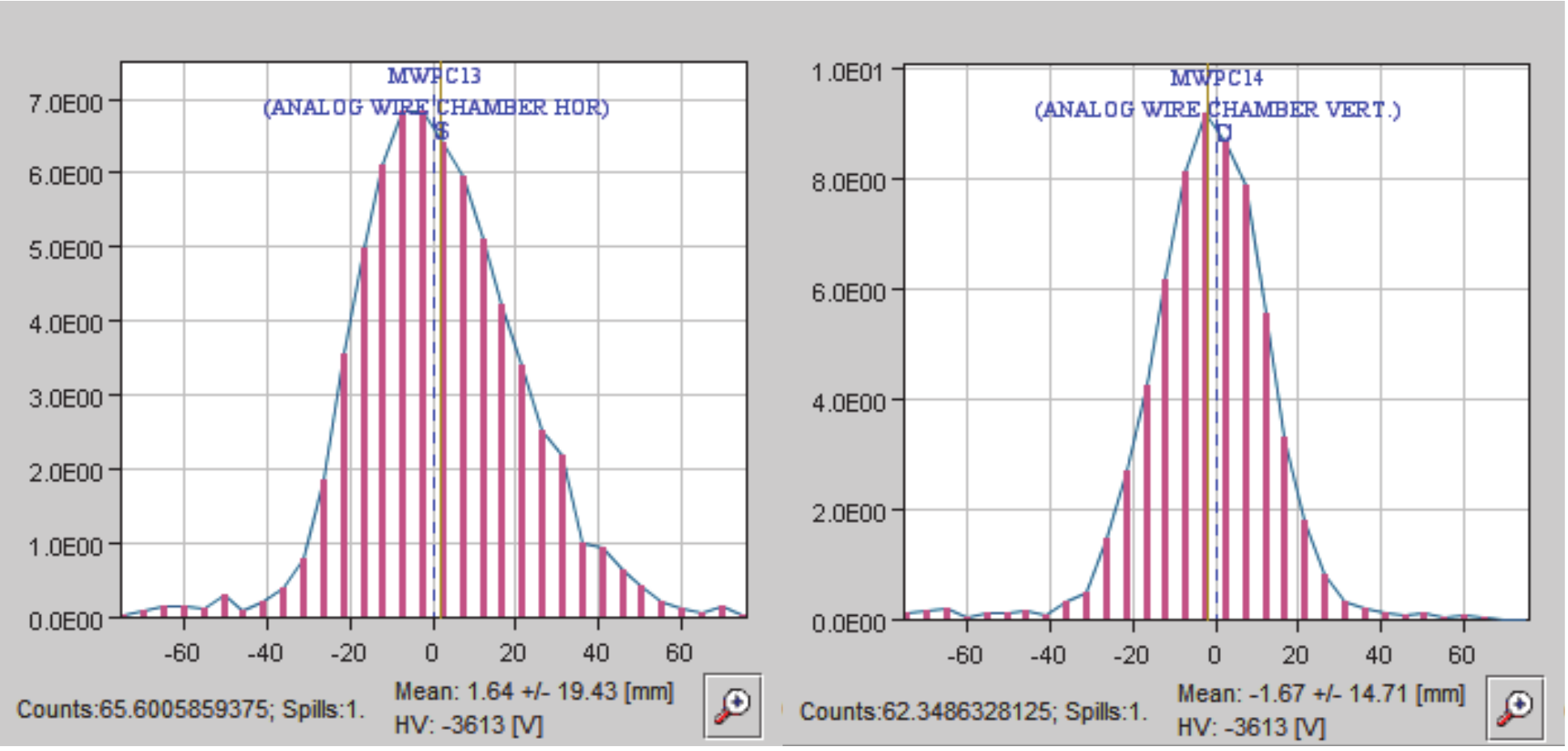}
\caption{Profile readouts from the first $20\times20$ cm$^2$ \textsc{gem} prototype in the M2 beamline.}
\label{M2}
\end{figure}

With energies of many GeV, these beams ionize the gas at least two orders of magnitude less than at the \textsc{ad}.
The gain required is therefore much higher, but multiple scattering and the angular distortions it causes are much less of a concern.
These constraints led us to a design of a triple \textsc{gem} detector of a more robust build, schematically pictured on the right of figure~\ref{SingleTriple}, and photographed in figure~\ref{BigGEM}.
The front and back of the detector is covered with stiff, light panels of a Rohacell structural foam\footnote{Rohacell XT. www.rohacell.com}, sandwiched between thin sheets of fiberglass.
These stiff panels contribute substantially to the material budget, which for these detectors is $0.85\% X_0$.
But the panels eliminate any issues with the entrance window and cathode bulging due to the gas pressure; and these prototypes are invulnerable to rough handling or repeated mechanical shock.
Also, the production of these panels, with integrated readout and high voltage circuitry and gas features, is very efficient in terms of time and cost, and leaves relatively little work to be done to assemble a chamber.
Moreover, all these detectors are motorized and can be moved out of the beam thus minimizing the importance of material budget.
Figure~\ref{M2} shows the first profiles measured with a $20\times20$ cm$^2$ prototype, at the M2 beam line.

We also tested $10\times10$ cm$^2$ triple \textsc{gem}s without the foam panels on the H4 and H6 beam lines.
They gave fine profiles consistent with the ones from the existing wire chambers.

\section{Conclusions and outlook}
We have studied the possibility to replace the \textsc{mwpc}s used as transverse profile monitors in different experimental areas at \textsc{cern} with \textsc{gem}-based detectors.
We built and tested single and triple \textsc{gem} detectors, for the low energy antiproton beam of the \textsc{ad} and the high energy beams in the experimental areas, respectively.
The \textsc{ad} needed replacement most urgently, and all their wire chambers were successfully replaced by well-performing single \textsc{gem}s.

The way we designed our \textsc{gem} detectors allows us to customize many parameters (number of \textsc{gem}s, width of gaps, robustness/material budget) during assembly.
We have thus produced a family of detectors from a 2D ionization chamber to a rugged $20\times20$ cm$^2$ triple \textsc{gem} detector.
These detectors all perform well, and are credible candidates to replace the \textsc{mwpc}s as profile monitors.

Compared to a typical \textsc{mwpc}, a typical triple \textsc{gem} detector represents roughly 3 times the material budget.
In many other aspects \textsc{gem} detectors exceed the performance of wire chambers.
Most notably, \textsc{gem} detectors feature probably the highest rate capability of any gaseous detector.
\textsc{Gem}s are also much more resilient to aging, particularly in Ar/CO$_2$ gas mixtures.
What makes \textsc{gem} detectors attractive from a practical point of view are a relatively low cost of production and operation, and the fact that they are essentially maintenance free.
The single-GEM detectors now installed at the AD are greatly appreciated for their simultaneous profile measurements of both planes allowing for reliable profile readout at low energy.
There are even more benefits to \textsc{gem} detectors that our studies have not at all explored, such as an excellent spatial resolution ($<50 \mu$m \cite{compass}) and time resolution ($<5$ ns \cite{lhcb}).
With a growing number of \textsc{gem} detectors used as beam profile monitors, both at \textsc{cern} and elsewhere, we expect this technology to become increasingly common in beam instrumentation.

\end{document}